# Gai Reply to Comment by Schumann *et al.*
# [https://arxiv.org/abs/1904.03023v1]

Abstract

Statements included in the comment published on the arXiv by Schumann *et al.* (https://arxiv.org/abs/1904.03023v1) are contradicted by documents that were communicated to one of the co-authors of the comment (Dr. Koester). These documents are reviewed but cannot be disclosed here due to copyright (they are available on request). A summary of the scientific dispute between the collaboration and Dr. Schumann, was submitted on September 24, 2018, to the Directorate Support of the Paul Scherrer Institute (PSI) and can be provided on request.

I serve as the spokes-person of the Israel-US-Switzerland collaboration representing: the SARAF, the Weizmann Institute, Bar-Ilan University, CERN, the PSI, and the University of Connecticut. Already on August 14, 2017, seven senior members of the collaboration signed off on a manuscript summarizing our results. We stated then: "**we agree [the paper] is now ready for publications and we request your [Dr. Schumann] comments within one week**". Since then the collaboration did not receive convincing critique of our paper and the discussion was steered away from the scientific content of our paper, as is the case here in this comment, which I must now address in my capacity as the spokes-person of the project.

In their comment dated April 3, 2019, on my posting in the arXiv [1] Schuman *et al.* [2] make a number of statements that were contradicted with documents that were communicated by myself on March 19, 2019, to a co-author of the comment (Dr. Koester) [2], as we discuss below.

1. Schumann *et al.* state: "The paper had been submitted by M. Gai as contribution to the proceedings of the NPA8 conference performed in June 2017 **without the knowledge of a considerable number of the coauthors listed in [27]**".

   In an email that I circulated on November 13, 2017, to all collaborators including all the authors of the comment [2], it was stated at the top: "**I attach the paper I submitted to NPA8**". Accordingly, a copy of my invited talk paper at the NPA8 was sent to all collaborators on November 13, 2017, only a few days after it was submitted to the organizers of the NPA8 on November 8, 2017.

2. Schumann et al. state: "M. Gai asserts that the paper was accepted for the proceedings of the named [NPA8] conference".

   The organizers of the NPA8 meeting informed me that my paper was reviewed by two referees and was accepted for publication, and as such it was listed on November 18, 2017, with the publisher (EPJ) among the accepted papers. The list of the to be published papers was due at the publisher on November 18, 2017.

3. Schumann *et al.* state: "[my paper] was retracted by M. Gai during the refereeing process".

   Very recently Dr. Koester revealed to the collaboration that he was asked to referee my NPA8 paper on November 13, 2017. Based on this Dr. Koester is assuming (incorrectly) that the refereeing process was still active. Again, the list of accepted papers (including my paper) was sent to the publisher (EPJ) on November 18, 2017. It now appears that the pending review of my paper by Dr. Koester, that was not received by November 18, 2017, could not and did not play a role in the Editor's decision to forward my paper to the publisher on November 18, 2017, when the papers were due at the publisher (EPJ).

4. Schumann *et al*. state: "M. Gai did never have the right to report on unpublished proprietary data of the entire collaboration".

   In the Appendix I include the post script note to the NPA8 paper as published in [1]. The post script note demonstrates that all collaborators including the authors of the comment [2], authorized (and in fact advertised) on February 4, 2017 my invited talk at the NPA8 meeting (not withstanding Dr. Schumann's change of mind that occurred five weeks later on March 14, 2017, as discussed in the post script note). In addition, all collaborators approved already a year before the NPA8 meeting a talk announcing our results in the DNP/APS meeting in Vancouver, Canada, on October 2016. Our data have been in the public domain for over a year before my invited talk at the NPA8 meeting.

5. During the Thanks Giving break on November 19 - 20, 2017, the collaboration met in Israel. After that meeting on November 20, 2017, I informed the Editors of the NPA8 conference that I offered to withdraw my paper that was accepted to the proceedings of the NPA8. An email informing the collaboration of my agreement to withdraw the paper, was circulated to all collaborators on

November 19, 2017, by the Israeli colleagues. I was forced to take this step since Dr. Schumann threatened not to show up to a planned collaboration meeting (the so called "Geneva Meeting" held on February 2, 2018). This was also a gesture of good will on my side with the hope that we can work out our disagreements. Alas, the Geneva meeting (and further discussions afterwards) failed and subsequently I decided to post on the arXiv [1] the very paper that was reviewed by two referees and were to be published a year earlier in the proceedings of the NPA8.

Over the last two years the collaboration received a number of communications signed by Dr. Schumann and Dr. Dressler. Starting from March 24, 2017 the collaboration replied with numerous emails contradicting the strong statements delivered by Dr. Schumann and Dr. Dressler. Several communications to Dr. Schumann were signed by all seven senior members of the Israel-US-Switzerland collaboration. On September 24, 2018, I summarized to the Directorate of the PSI the continuous disagreements the collaboration had with Dr. Schumann and Dr. Dressler; this document is not confidential and is available on request.

In a scientific discourse it is customary for disagreeing scientists to publish separate papers detailing their disagreements. However, Dr. Schumann chose instead to block our publications. In fact, Dr. Schumann invoked "veto rights" that were never stipulated or agreed by the collaboration and contradict the very essence of the scientific process, no less it contradict section 734.8(a) of the Export Administration Regulations (EAR) of the USA [3].

Numerous communications to conference organizers in Italy (NPA8 and the Santa Tecla meeting), in Cuba (Latin American School) and in Mexico (the Cocoyoc meeting), were sent by Dr. Schumann with a demand to cancel my invited talks and/or not to publish the paper of my invited talk. Encouraged by Dr. Schumann, the administration of the PSI sent numerous strongly worded communications to the Dean at the University of Connecticut and in one communication they accused myself, the spokes-person of this international collaboration, of committing "continuous misdemeanor" (i.e. a crime punishable by up to one year in jail or $1,000 fine).

A scientific debate in peer reviewed publications is the essential pillar upon which the scientific process rests. And it will be followed here, as it should be. This comment is only intended to correct the statements made in [2].

I append below the Post Script note included in my posting on the arXiv [1]:

4 Post Script Remark

The current paper presents the results of the SARAF US-Israel-Switzerland collaboration that were approved for public presentation in the DNP meeting of the American Physical Society in Vancouver, Canada on October 14, 2016 [19]. All members of the collaboration listed as co-authors in [19] approved the results presented in the DNP meeting in October 2016. In addition this invited talk by the author and a second poster paper by E.E. Kading et al. in this NPA8 meeting [27], was approved by the collaboration on February 2, 2017. Specifically all co-authors listed the NPA8 poster paper [27] approved on February 4, 2017 the paper [27] which also referenced the invited talk of the author. On March 14, 2017, the collaboration learned that two colleagues had a change of mind. Their claims were seriously considered by the collaboration, some changes were adopted, but the essential claims made in the communication of March 14, 2017, were refuted by the rest of the collaboration on March 24, 2017. While the discussion is still going on, the collaboration did not as of yet received convincing argument that invalidates the material that was approved for public announcement already a year ago in October 2016 and again for the second time in this NPA8 meeting. The collaboration is committed to continue the internal scientific dialogue and we will continue to judge statements of facts based on their scientific merit.